# Aperiodicity is all you need: Aperiodic monotiles for high-performance composites

Jiyoung Jung[1,2], Ailin Chen[1,2], and Grace X. Gu[1,*]

[1]Department of Mechanical Engineering, University of California, Berkeley, CA 94720, USA
[2]These authors contributed equally to this work
[*]Corresponding author: ggu@berkeley.edu

**Abstract**

This study introduces a novel approach to composite design by employing aperiodic monotiles, shapes that cover surfaces without translational symmetry. Using a combined computational and experimental approach, we study the fracture behavior of composites crafted with these monotiles, and compared their performance against conventional honeycomb patterns. Remarkably, our aperiodic monotile-based composites exhibited superior stiffness, strength, and toughness in comparison to honeycomb designs. This study suggests that leveraging the inherent disorder of aperiodic structures can usher in a new generation of robust and resilient materials.

## 1. Introduction

Composite materials, celebrated for their customizable mechanical properties, serve as lightweight structural components that are integral in aerospace and biomedical sectors.[1-5] The strength of these materials lies in their composite nature – combining properties of different base materials allows the creation of a composite with a harmonious balance of multiple desired properties. This concept is beautifully exemplified in biological materials[6-11] such as nacre and wood, which generally outperform their engineering counterparts in mechanical performance, despite being composed of relatively weak constituents. Traditional engineering composites are often characterized by repeating unit cells, a feature that simplifies the design and manufacturing processes. However, such ordered structures can lead to catastrophic failure under critical loading. Meanwhile, biological materials often present disordered structures, where the unit cells vary spatially.[12] The extent to which this disorder plays a role in the improved mechanical performance of biological materials remains a topic of ongoing research.

The inherent benefits of materials with irregular or disordered microstructures have recently garnered scientific interest.[13-15] Characterized by heterogeneous microstructures, these structures could offer a fortified path for stress wave propagation, thereby increasing resilience under heavy loads.[16-19] Emerging research indicates that by amplifying this irregularity, the flaw tolerance of specific cellular frameworks can be enhanced.[20] Moreover, the microscopic intricacies of polycrystalline configurations, encompassing grain boundaries, precipitates, and phases, are perceived as prospective templates for engineering materials with enhanced toughness.[21,22] Current methodologies for creating these heterostructures involve techniques such as randomly moving nodes within regular lattice structures, constructing material foams, or stacking materials with different microstructures[17,23,24] However, these methods introduces a layer of complexity to design and manufacturing, especially with challenges due to the imperfect assembly of differently oriented unit cells.

Addressing these challenges, our study presents the integration of aperiodic monotiles in composite designs. Aperiodic monotiles, as discovered in recent literature, have been shown to cover a surface entirely with intrinsic aperiodicity.[25] This makes them an ideal choice for creating disordered materials. The usage of aperiodic monotiles in composite design would facilitate tunable properties while maintaining excellent interface bonding. In this work, we explore a completely new family of architecture composed of aperiodic monotiles for creating composite materials. Specifically, we developed a numerical phase-field model to simulate the properties and crack propagation of composites consisting of aperiodic monotiles. Our models are validated with tensile experiments of additively manufactured specimens. The aperiodic monotile-based designs are benchmarked with periodic honeycomb-based design, which is one of the most widely used shapes in engineering applications due to its superior mechanical performance.[26] It is envisioned that these types of aperiodic designs could lead to the development of stronger and tougher composite materials compared to the conventional periodic designs.

## 2. Results and discussions

An aperiodic monotile is a shape that can cover a two-dimensional (2D) surface without any translational symmetry or a repeating pattern.[25] An example schematic of tiling using a 'hat' polykite aperiodic monotile[25] is shown in Fig. 1 (a). Due to the characteristics of the hat monotile generated based on the hexagonal structure, the monotile can have six rotational angles and a flipped shape as shown in Fig. 1 (b, c). These hat monotiles covers the infinite plane in an irregular manner, which enables limitless designs by translation and rotation of tiles. Here, we will introduce translation and rotation of the aperiodic monotile-based design along the reference tile, part of the infinite reference tile shown in Fig. 1 (a), and study their influence on mechanical behavior.

Experiments are conducted to explore the mechanical performance of the aperiodic monotile-based composites. Mechanical tensile tests are conducted, with more details in the Methods section. In terms of material selection, two base constituents are utilized, one for the boundaries and another for the inner unit cell material, affording us flexibility in generating composites with a range of mechanical properties. To evaluate the composites under realistic operational conditions, a defect is introduced, in the form of a notch, into the samples. This approach enables us to investigate the tolerance of these materials to such anomalies. For the fabrication process, Polyjet additive manufacturing (Stratasys Connex 3) is employed, utilizing digital photopolymer materials with a modulus range spanning over three orders of magnitude. Two digital photopolymer materials are used to achieve the required characteristics. In this case, TangoBlackPlus, the softer material, is used to form the boundaries, while VeroClear, the stiffer material, is used in the core areas. A specimen has dimensions of 50 mm by 125 mm by 3 mm with a notch of 20% length of the specimen width (10 mm) as shown in Fig. 2 (a). We prepared the specimens with different notch tip locations and volume fractions. For aperiodic monotile-based specimens, we have prepared three different specimens with a volume fraction of 80% VeroClear and 20% TangoBlackPlus where two specimens are subjected to a planar translation (denoted as AP80_T1 and AP80_T2) and a third specimen undergoes a rotation of 30 degrees (denoted as AP80_R1) along the infinite reference tile. Additionally, we also test two specimens with a volume fraction of 70% VeroClear and 30% TangoBlackPlus with translation and rotation (denoted as AP70_T1 and AP70_R1). These designs are shown in Fig. 2 (a). We note that infinitely many patterns can be generated by the translation or rotation of the tiling. As a benchmark, the honeycomb structure which has a periodic pattern is considered as shown in Fig. 2 (b). Two honeycomb-based specimens with different volume fractions are considered, denoted as HC80 and HC70.

Experimental stress-strain curve results for the various samples are presented in Fig. 3 (a) and (b), where (a) and (b) are results for 80% and 70% volume fractions of VeroClear material, respectively. The shadows in the curves represent variations from three experiments for each sample design. From the curves, it can be seen that the aperiodic monotile based structures show higher stiffness, strength and toughness compared to the honeycomb structures for both volume fraction cases. In Fig. 3 (c), the crack propagation paths for different designs are displayed. The honeycomb structure exhibits a path reminiscent of brittle fracture, whereas the aperiodic monotile-based structures reveal a multifaceted crack trajectory with a combination of large and small zigzags, enhancing crack resistance. The stiffness, strength, and toughness values are compared in Fig. 3 (d). With an 80% volume fraction of VeroClear material, the aperiodic structure is 103% superior in stiffness, 34.5% in strength, and 15.9% in toughness compared to the honeycomb structure. These results indicate that the aperiodic monotile-based composites have not only higher stiffness but also higher strength and toughness compared to the honeycomb structure showing the mechanical superiority of the aperiodic structure, which can be advantageous in many applications.

To probe further into the mechanisms, simulations utilizing phase-field modeling are performed for the aperiodic monotile-based design and the honeycomb-based design. More details about the phase-field model are discussed in the Methods section. The simulations are carried out on 2D specimens, each measuring 50 mm by 75 mm (without gripping section) and featuring a 20% length crack of sample, under the tensile loading. The simulations have utilized material properties obtained from the characterization of the base materials shown in **Table 1**. The stress-strain curves for the various designs are shown in Fig. 4, with the general trends matching the experimental results. Here, the aperiodic monotile-based structures show higher stiffness, strength, and toughness compared to the honeycomb structures. Additionally, it can be seen that the aperiodic monotile based structure show relatively consistent mechanical performance (similar to experiments) regardless of the crack location determined by the translation or rotation of the tiling. This points to the potential defect tolerance capabilities of these types of aperiodic structures. The crack propagation behavior of the aperiodic monotile specimen (AP80_T2 in Fig. 4 (a)) with different strains can be found in Fig. 5 (a, c). The crack propagation behavior of the honeycomb structure with different strains can be found in Fig. 5 (b, d). In this model, a phase value of 1 symbolizes complete damage (represented by the color red), while a phase value of 0 denotes no damage (represented by the color blue). Elements exhibiting more than 98% damage are not depicted. In the phase-field model, as the strain increases, the phase value of the element under stress increases, which indicates the degree of damage and crack propagation. From Fig. 5 (a, c), the aperiodic monotile structures show a complex crack path mixed with large and small zigzags, behavior that is also seen in experiments. Conversely, the crack path in the honeycomb specimen pursued the shortest possible trajectory as shown in Fig. 5 (b, d). These results indicate that phase-field modeling holds the potential for capturing the fracture behaviors of these unique composite systems, hence offering promise for future exploration.

**Conclusions**

This study introduced new architectures incorporating aperiodic monotiles into composite designs. Not only do these structures greatly simplify design and manufacturing, but their aperiodicity also offers a promising path to enhanced mechanical resilience. Tensile experiments and corresponding numerical phase-field models show that these aperiodic monotile designs outperform traditional honeycomb-based designs in terms of stiffness, strength, and toughness. Furthermore, our findings highlighted the aperiodic designs' inherent capability to tolerate defects. Through the synthesis of aperiodic materials design, advanced manufacturing techniques, and numerical simulations, this research illuminates a promising avenue for the next generation of composite materials.

**Methods**

**Mechanical testing:** Experimental tensile tests (Mode I fracture) on both aperiodic monotile and honeycomb structures are conducted. To secure the specimens, mechanical vise action grips are utilized, clamping only the designated gripping area made of the VeroClear material. The tests are conducted at a controlled tensile displacement rate of 2 mm/min. Throughout the tests, force, displacement, and time data are recorded at a frequency of 1.04 Hz. The test terminates when the force is dropped to nearly zero, and the crack fully propagated through the transverse direction of the specimen. At least three specimens are printed and tested for each microstructure.

**Phase-field modeling:** Phase-field modeling is employed due to its established capabilities in simulating intricate crack evolution phenomena including curvilinear crack paths and crack branching.[27-29] Among various versions of phase-field modeling, we adopt a hybrid formulation-based phase modeling that can be applicable for combined shear and tensile loading,[30] which enables modeling crack propagation for composites with complex microstructures. The phase-field modeling is conducted using ABAQUS with user-defined element (UEL) subroutine. The model consisted of three layers including the phase-field layer, displacement layer, and visualization layer sharing the identical nodes. Approximately 150,000 quadrilateral plane stress elements are used for each layer of specimens. The y-directional displacement of the lower surface is fixed, and the displacement control is applied to the upper surface.

**Tables**

Table 1. Material properties of VeroClear (stiff) and TangoBlackPlus (soft) for phase-field model which were fitted based on experimental characterization results.

|  | VeroClear | TangoBlackPlus |
|---|---|---|
| Young's modulus ($E$) [MPa] | 1000 | 0.6 |
| Poisson's ratio ($\nu$) | 0.35 | 0.49 |
| Critical energy release rate ($g_c$) [kN/mm] | 3.0 | 0.3 |
| Regularization parameter ($l$) [mm] | 0.4 | 0.4 |

**Figures and captions**

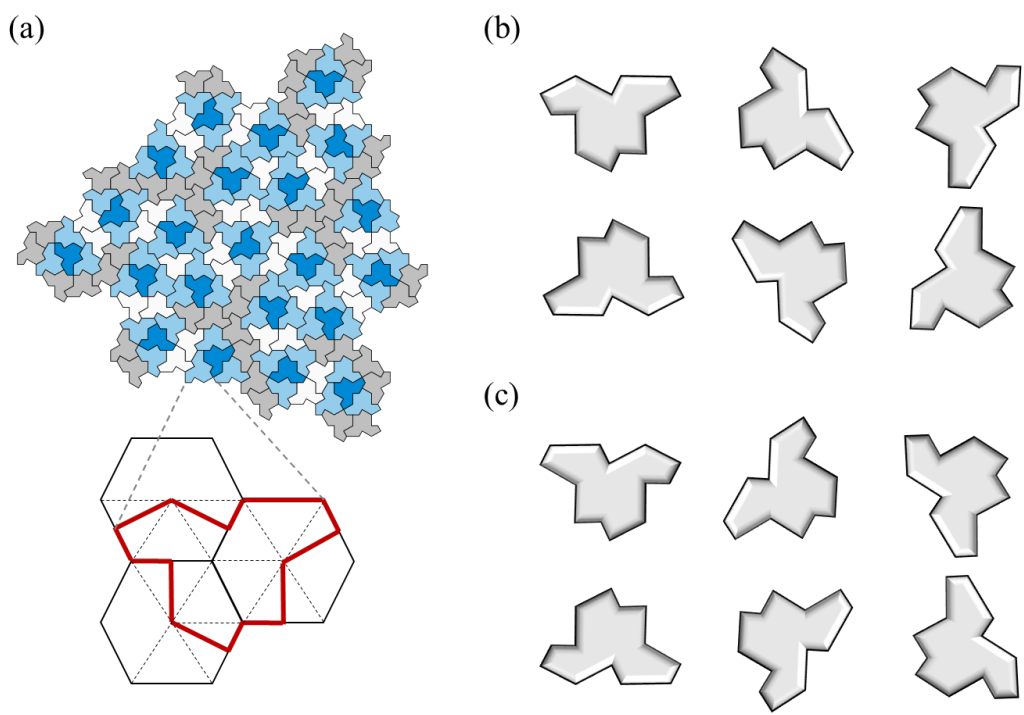

**Figure 1**. (a) Tiling schematic of a hat (polykite) aperiodic monotile. Hat monotiles can have (b) six rotational angles and (c) a flipped shape of them. The flipped tiles are presented as dark blue color in (a).

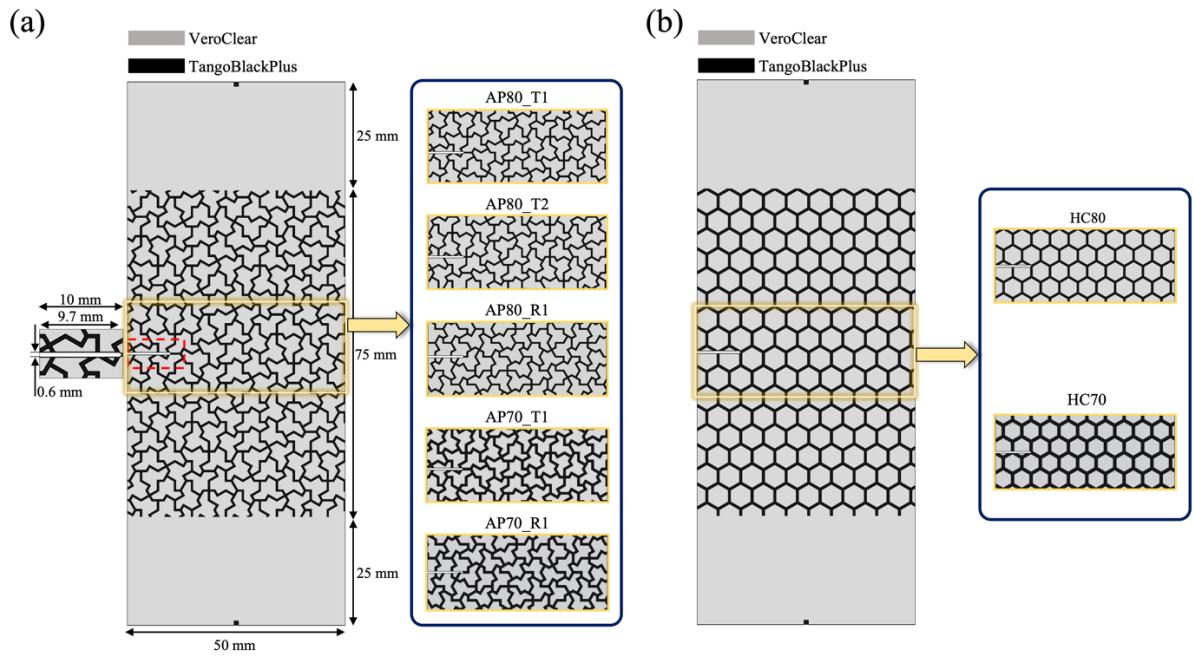

**Figure 2**. Specimen schematics for aperiodic monotile and honeycomb composite structures. (a) Designs of aperiodic monotiles based structures with different locations, angles, and volume fractions (80% and 70% volume fractions of VeroClear material) are presented. (b) Designs of honeycomb structures with different volume fractions (80% and 70% volume fractions of VeroClear material) are presented. The thickness of the specimens is 3 mm.

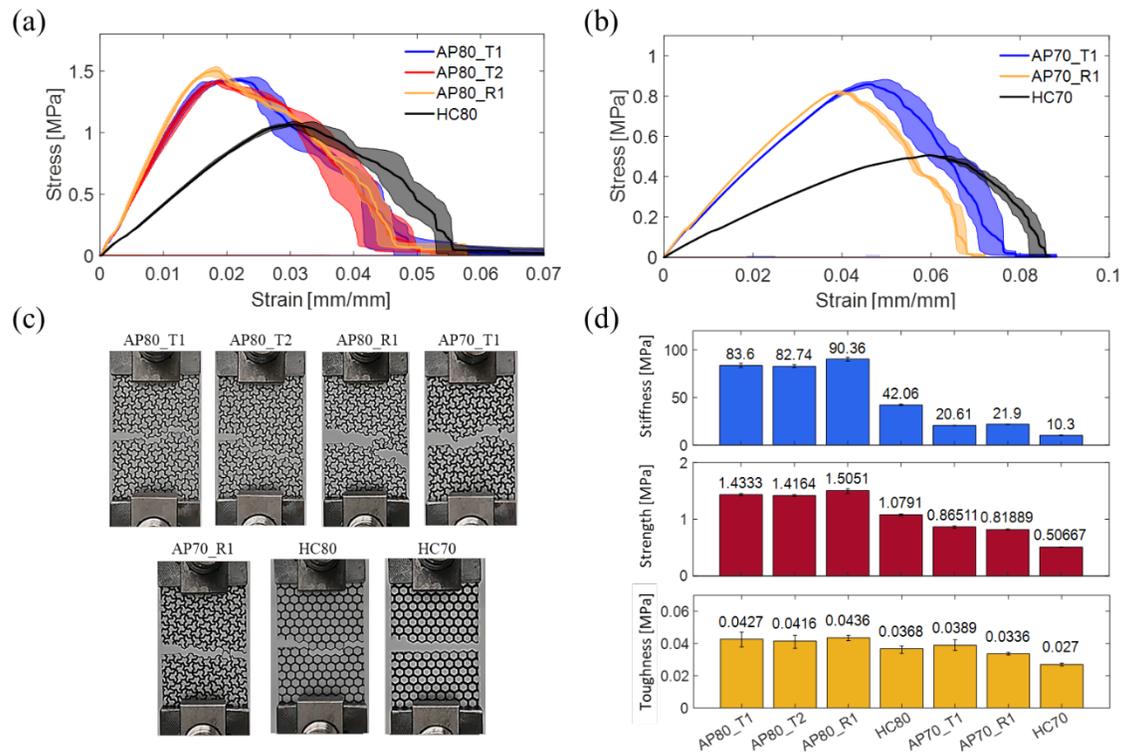

**Figure 3.** Experimental results for aperiodic monotile (AP) and honeycomb structures (HC) for (a) 80% and (b) 70% volume fractions of VeroClear material. Solid lines represent one of the three repeated experiments, and the variation is presented as a shaded area. (c) The damage and crack propagation path for the various specimens after experiments. (d) Comparison of stiffness, strength, and toughness values for the various structures.

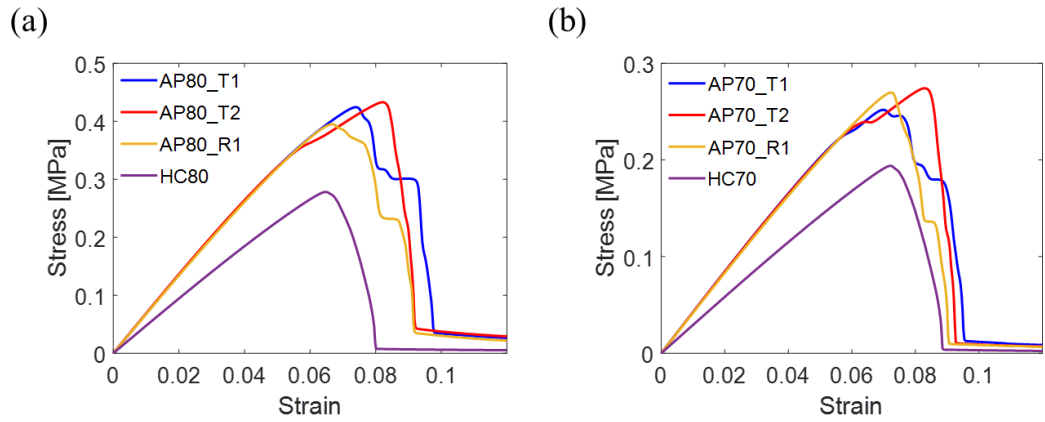

**Figure 4**. Finite element analysis results using phase-field model for aperiodic monotile (AP) and honeycomb (HC) structures with (a) 80% and (b) 70% volume fractions of VeroClear material. For aperiodic monotile based structures, two translated specimens (T1 and T2) and one rotated specimen (R1) are examined, and the honeycomb structures are considered as a benchmark comparison.

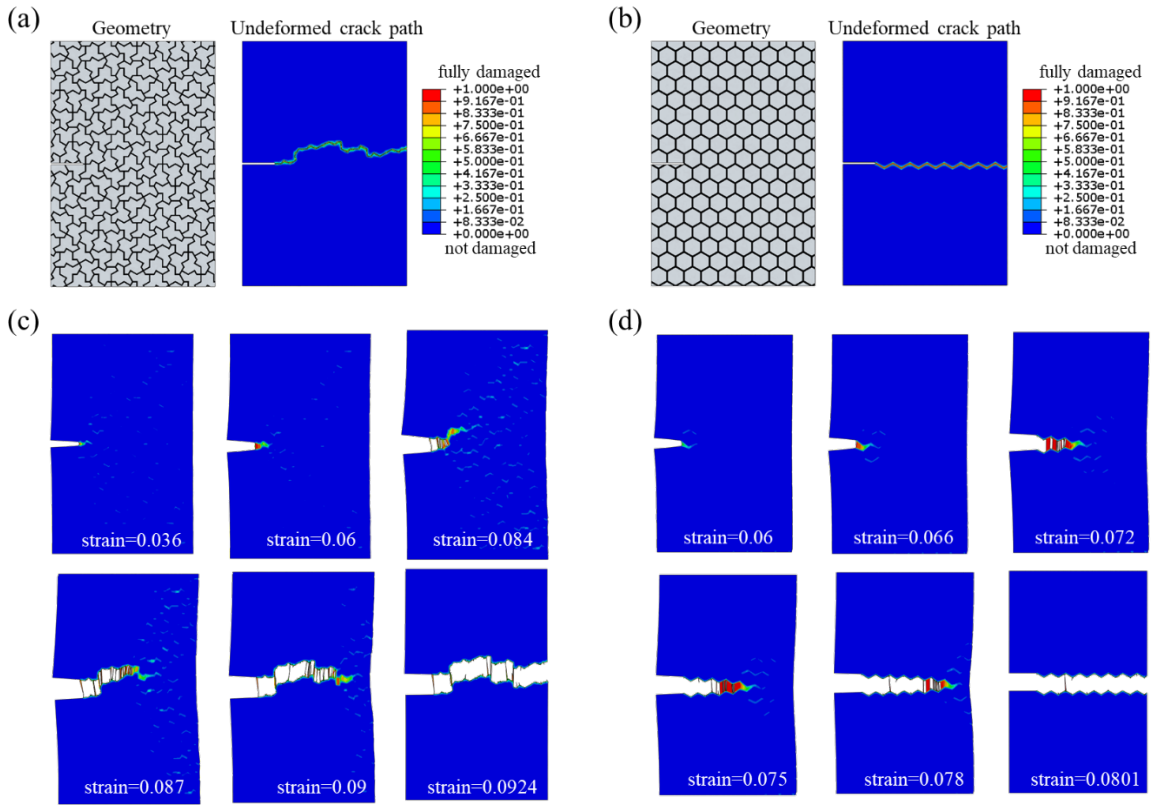

**Figure 5**. Comparison results from the phase-field model for aperiodic monotile design (AP80_T2) and honeycomb structure (HC80). (a, b) Geometry and undeformed crack propagation results of the specimen are presented. Crack propagation behavior with strain increments (scale factor =1) for (c) the aperiodic monotile structure and (d) the honeycomb structure, respectively. Elements with over 98% damage are not shown.